\documentclass [preprint,aps,showpacs]{revtex4}
\usepackage[final]{graphics}
\usepackage{amssymb}
\usepackage{amsfonts}
\usepackage{epsfig}
\usepackage{graphicx}

\begin{document}
%\draft
\title{Particle redistribution and slow decay of correlations in
hard-core fluids on a half-driven ladder}
\author{Ronald Dickman\footnote{dickman@fisica.ufmg.br} and
Ronaldo R. Vidigal\footnote{rvidigal@dedalus.lcc.ufmg.br}}
\address{Departamento de F\'{\i}sica, Instituto de Ci\^encias Exatas,\\
Universidade Federal de Minas Gerais \\
C. P. 702, 30123-970, Belo Horizonte, Minas Gerais - Brazil }

\date{\today}

\begin{abstract}
We study driven particle systems with excluded volume interactions
on a two-lane ladder with periodic boundaries, using Monte Carlo
simulation, cluster mean-field theory, and numerical
solution of the master equation. Particles in one lane are subject
to a drive that forbids motion along one direction, while in the
other lane the motion is unbiased; particles may jump between lanes.
Despite the symmetry of the rates for transitions between lanes, the
associated particle densities are unequal: at low densities there is
an excess of particles in the undriven lane, while at higher
densities the tendency is reversed. Similar results are found for an
off-lattice model.  We quantify the reduction in the stationary
entropy caused by the drive.  The stationary two-point correlation
functions are found to decay algebraically, both on- and
off-lattice.   In the latter case the exponent governing the decay
varies continuously with the density.
\end{abstract}

\pacs{05.40.-a, 05.10.-a, 02.50.Ga, 05.50.+q}

\maketitle

\section{Introduction}

Over the last decades considerable effort has been devoted to the
investigation of simple nonequilibrium models of interacting
particles. In addition to their application to specific problems
(traffic flow, ionic conductors, epidemics, etc.), such studies have
been motivated by the hope of developing a macroscopic description
of far-from-equilibrium systems. Driven lattice gases, also known as
{\em driven diffusive systems} (DDS) \cite{zia,schutz,marro}, have
played an important role in this program. In these models, a
stochastic lattice gas is subject to a ``drive" that biases hopping
along one of the principal axes of the lattice \cite{kat83}. The
repulsive version \cite{leu89,dic90,szabo94,szabo96} serves as a
model for fast ionic conductors \cite{boy79}. Stationary properties
of driven lattice gases depend strongly on the details of dynamics,
unlike in equilibrium. DDS have provided a wealth of unexpected and
counterintuitive results, many of which remain to be fully explained
\cite{zia}. (For example: the fact that the critical temperature of
the attractive lattice gas {\it increases} under driving.)

Recently, a study of a two-dimensional lattice gas with
nearest-neighbor exclusion (NNE), subject to a nonuniform
(shear-like) drive, revealed a tendency for particles to migrate to
the region with weaker drive \cite{nne-shear}; the accumulation of
particles is sufficiently strong to provoke sublattice ordering in
this region. At higher densities the trend reverses, with particles
accumulating in jammed regions that form where the drive is
strongest.  To better understand these results, we consider, in the
present work, a one-dimensional (ladder) analog of the model studied
in Ref. \cite{nne-shear}. The reduced dimensionality permits us to
obtain simulation results of high precision, to develop high-order
cluster approximations, and to derive essentially exact results (for
smaller systems), via numerical solution of the master equation. We
find that particles migrate preferentially to the undriven lane at
low densities, and to the driven one at higher densities. The same
tendencies are observed in simulations of a continuous-space model.

In our model, one ring corresponds, in isolation, to a totally
asymmetric exclusion process (TASEP) with sequential dynamics, while
the other is a symmetric exclusion process. (Note however that the
most widely studied version of the TASEP features on-site exclusion
only.) Both models are exactly soluble and possess a stationary
probability distribution that is uniform on the set of allowed
configurations.  The interaction between particles on different
rings and the possibility of jumping between them changes this
situation: the stationary probability distribution is no longer
uniform, and the model exhibits nontrivial behavior. (The ladder
without driving in either lane again corresponds to a simple
equilibrium model that can be solved exactly via the transfer-matrix
technique.) The TASEP has attracted much attention due to the phase
transitions that arise when, instead of periodic boundaries, the end
sites are coupled to particle reservoirs
\cite{derrida98,schutz,krug,derrida,kolomeisky}. Although not
directly to the system studied here, it is worth noting that quite
surprising phase behavior is found in an interacting TASEP ladder,
when the two lanes are coupled to different reservoirs
\cite{popkov}, or there are asymmetric transition rates between them
\cite{pronina}.  A single-lane TASEP with repulsive nearest-neighbor
interactions exhibits a surprisingly complex phase diagram
\cite{popkov99}.

Our focus in this paper is on stationary properties, specifically,
the particle density in the two lanes, the current, and the
two-point correlation function.  We also examine the stationary
probability distribution (for the lattice model) on configuration
space, and show that this distribution does not admit a simple
characterization in terms of a set of macroscopic variables. The
two-point correlation function $h(r)$ exhibits power-law decay in
both the lattice and continuous-space models.  Algebraic decay of
correlations is well established in DDS \cite{zia,zhang88}, and is
expected to be a generic feature of anisotropic nonequilibrium
systems with a conserved density \cite{garrido90}, but is not
commonly observed in periodic, one-dimensional systems.

The balance of this paper is organized as follows.  In the next
section we define the models, and in Sec. III discuss the stationary
solution of the master equation for the lattice gas model. Cluster
approximations are developed in Sec. IV.  In Sec. V we report
simulation results, and in Sec. VI present a qualitative explanation
of the migration observed at low density. We close in Sec. VII with
a discussion of our findings.

\section{Model systems}

We consider particle systems with excluded-volume interactions on
ladders with periodic boundaries.  The position of particle $i$ is
$(x_i,y_i)$, where $y_i \in \{1, 2\}$ denotes the {\it lane}. In the
lattice gas model, $x_i \in \{1, 2, ..., L\}$; in the off-lattice
model $x_i \in [0, L]$.  In both cases, there are periodic
boundaries in the $x$ direction.  Interactions between particles are
purely repulsive: in the lattice gas model, the minimum allowed
distance between particles in the same lane is 2, while for
particles in different lanes, the minimum allowed separation is 1.
(Note that the NNE condition means there is no particle-hole
syttetry.)

In the {\it off-lattice model} the minimum distance between
particles in the same lane is 1, while the minimum separation along
$x$, for particles in different lanes is $1/2$. The maximum density
is 1 in this case. In both the lattice gas and continuous-space
models, the excluded volume constraint makes it impossible for
particles to pass one another: the order of particles along the $x$
direction cannot change.

\begin{figure}[h]
\rotatebox{0}{\epsfig{file=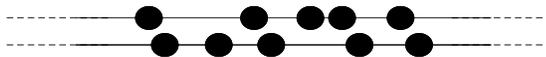,width=7.0cm,height=4.0cm}}
\caption{Typical region in the continuous-space model.  Particles in
the upper lane may only move to the right.} \label{conf}
\end{figure}

The process evolves via a sequential (continuous-time) Markovian
dynamics. In the lattice gas, each particle in the driven lane
($y=2$) attempts to hop forward by one lattice spacing [$(x,2) \to
(x+1, 2)$] at rate $\alpha$, and to hop to the other lane [$(x,2)
\to (x,1)$] at rate $\beta$. Particles in the undriven lane ($y=1$)
attempt to hop to the upper lane at rate $\beta$, and to hop forward
or backward, each at rate $\gamma$. Any attempted move that does not
violate the excluded-volume constraint is accepted. The
continuous-space model follows a similar dynamics. The rate of
interlane driven transitions is again $\beta$.  In the driven lane,
the intralane hopping rate is $\alpha$, with the displacement
$\Delta x$ uniformly distributed on $[0,1]$, so that particles in
this lane cannot jump in the $-x$ direction. The hopping rate within
the undriven lane is $\gamma$; the displacement is uniformly
distributed on $[-1/2, 1/2]$. Fig. \ref{conf} shows a typical region
in the continuous-space model.
\vspace{-5em}

The parameters characterizing the system are its size $L$, particle
number $N$, and the transition rates $\alpha$, $\beta$, and
$\gamma$.  We expect that for large $L$, macroscopic properties
depend on the particle density $\rho = N/(2L)$ instead of on $N$ and
$L$ separately.  In general, stationary properties depend on the
transition-rate ratios ($\tilde{\beta} \equiv \beta/\alpha$ and
$\tilde{\gamma} \equiv \gamma/\alpha$, for example). The models (on-
or off-lattice) with unbiased dynamics in both lanes represent {\it
equilibrium} fluids with purely excluded volume interactions. In
this case the stationary probability distribution is uniform on the
set of allowed configurations.  The NNE lattice gas with both lanes
driven also possesses a uniform stationary distribution, provided
any absorbing configurations (see below) are excluded from
consideration.  (To show this one verifies that the number of
configurations accessible from a given configuration ${\cal C}$
equals the number of configurations from which ${\cal C}$ is
accessible, and that each transition out of ${\cal C}$ occurs at the
same rate as the corresponding transition in.  It is easy to verify
that this symmetry no longer holds in the half-driven ladder.) The
same holds for the half-driven ladder with on-site exclusion only.
We should therefore expect these models to have a trivial behavior
(for example, exponentially decaying correlations).

\section{Master equation}

It is straightforward to write the master equation for the lattice
gas model defined in the previous section, the only practical
limitation coming from the rapid growth in the number of
configurations and transitions with increasing system size.  In both
this and the following section we employ a base-3 notation in which
the configuration $c$ in each column (fixed $x$) is denoted by 0, 1
or 2, corresponding to both sites empty, lower site occupied, or
upper site occupied, respectively.  Then a configuration on a ring
of $2 \times L$ sites is denoted by ${\cal C} = \sum_{j=1}^L 3^{j-1}
c_j $.  Allowed configurations are such that, if $c_j = 1$ or 2,
$c_{j-1} \neq c_j$ and $c_{j+1} \neq c_j$, for $j = 1$,...,$L$, with
periodic boundaries.

Our analysis of the master equation involves three steps. First, we
enumerate all allowed configurations of $N$ particles on a ladder of
$2L$ sites; configurations that differ merely by a rotation are
treated as equivalent.  Next, the set of all transitions and
associated rates is generating by considering all possible particle
movements in each configuration. Finally, the stationary probability
distribution is generated via the fast iterative method described in
\cite{intme}. (For an alternative approach to constructing the
stationary distribution see \cite{zia06}.)

At higher densities, some of the allowed configurations are {\it
absorbing}; all particle motion is blocked.  This is obviously the
case for $N=L$, which is of no interest here.  For even $L$,
however, absorbing configurations with $N < L$ are possible;
instances appear for $N$ as small as $2L/3$, if $L$ is a multiple of
six. (In particular, all configurations with $N=L-1$ are absorbing
for $L$ even.) It turns out that all absorbing configurations are
isolated in the sense that there are no transitions into them from
other configurations.  We therefore eliminate such configurations
from the state space.  (In simulations the initial configuration is
always chosen to be nonabsorbing.)  We have verified, for systems of
up to 38 sites, that the process (restricted, if necessary, to the
nonabsorbing subspace) is ergodic, i.e., there is a sequence of
transitions leading from a given configuration to any other.

\begin{figure}[h]
\rotatebox{0}{\epsfig{file=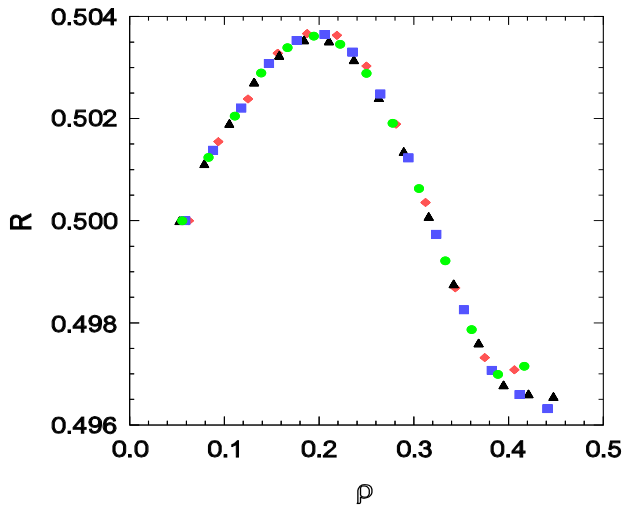,width=7.0cm,height=7.0cm}}
\caption{Stationary fraction $R$ of particles in the undriven lane
versus total density $\rho$.  Diamonds: $L=16$; squares: $L=17$;
circles: $L=18$; triangles: $L=19$.} \label{sd1619r}
\end{figure}

The stationary probability distribution $\overline{p}_i$ is used to
evaluate the particle and current densities in each lane, and the
statistical entropy per site, $s = - \frac{1}{2L} \sum_i
\overline{p}_i \ln \overline{p}_i$.  We report results for the case
$\alpha=2$, $\beta= \gamma = 1$.  Fig. \ref{sd1619r} shows the
fraction $R = \rho_1/\rho$ of particles in the nondriven lane
($y=1$) for several system sizes. For densities less than about 0.3,
there is an excess of particles in the undriven lane; the excess
appears for systems with three or more particles. For densities
above 0.3, the trend reverses, with an excess of particles appearing
in the driven lane. Away from the extremes of $\rho \simeq 0$ and
$\rho \simeq 1/2$, the fraction $R(\rho)$ appears to trace out a
smooth, roughly sinusoidal curve.  (Note however that this curve is
not symmetric about the line $\rho=1/4$, nor about the line
$R=1/2$.) We verified that in the absence of driving, the stationary
probability distribution is uniform on the set of allowed
configurations.

\begin{figure}[h]
\rotatebox{0}{\epsfig{file=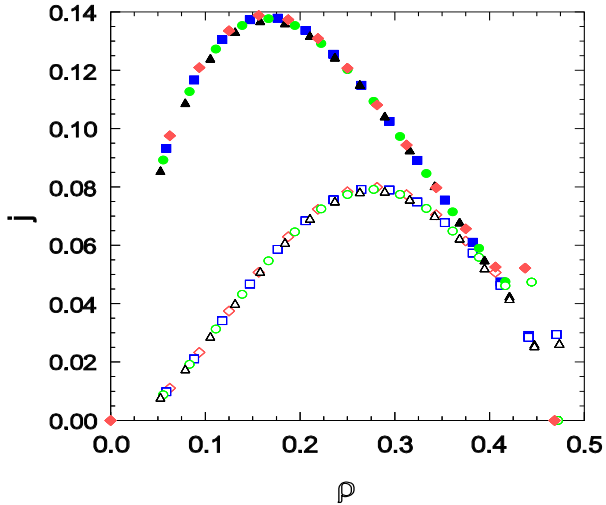,width=7.0cm,height=7.0cm}}
\caption{Stationary current density $j$ versus total density $\rho$.
System sizes as in Fig. \ref{sd1619r}.  Upper curve: driven lane;
lower: undriven lane.} \label{sd1619j}
\end{figure}

In Fig. \ref{sd1619j} we show the current densities  (mean rate of
moves in the $+x$ direction, less the number in the $-x$ direction,
divided by $L$), $j_d$ and $j_u$, in the driven and undriven lanes,
respectively. As expected, the current in the driven lane first
grows with $\rho$ (due to increasing number of carriers), and then
decays (due to blocking of particle movement at higher densities);
the maximum occurs at a density near 0.16. Due to the no-passing
condition, particles in the undriven lane must also, on average,
move forward. The current in this lane grows more slowly, reaching a
maximum near $\rho \simeq 0.28$.  At higher densities the currents
in the two lanes are virtually identical, as movement in one implies
a corresponding movement in the other.

The ``outlying" points at high density in Fig. \ref{sd1619j} reflect
an odd-even effect that occurs near maximum filling.  It is easy to
verify, for example, that an isolated hole ($N=L-1$) is immobile for
$L$ even, while for $L$ odd it has a stationary velocity of $2
\alpha \gamma/(\alpha + 2 \gamma)$. Surprisingly, when $L$ is odd,
the current for two holes is slightly {\it less} than for just one
hole.  These ``anomalies" are also observed in simulations (Sec. V).

\begin{figure}[h]
\rotatebox{0}{\epsfig{file=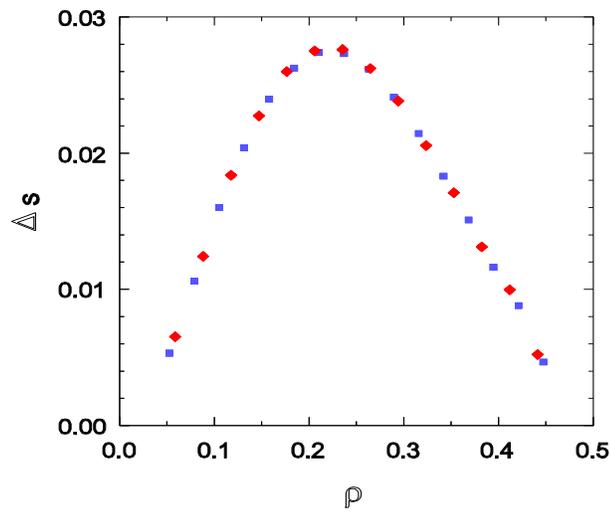,width=7.0cm,height=7.0cm}}
\caption{Entropy difference per site $\Delta s$ for $L=17$
(diamonds) and $L=19$ (squares).} \label{dels}
\end{figure}

\begin{figure}[h]
\rotatebox{0}{\epsfig{file=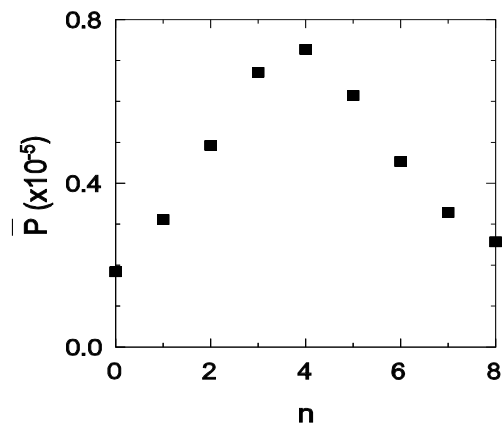,width=7.0cm,height=7.0cm}}
\caption{Average stationary probability of configurations having
exactly $n$ particles in the driven lane, $L=19$, $N=8$.}
\label{sdn198}
\end{figure}

As noted, in equilibrium all allowed configurations have the same
probability, so that the entropy (in units of Boltzmann's constant)
is $\ln \Omega(L,n)$, with $\Omega(L,n)$ denoting the number of
configurations of $n$ particles on a ladder of $2L$ sites that
satisfy the NNE condition. Under biased dynamics the stationary
probability distribution is no longer uniform over the set of
configurations, making the statistical entropy of the driven system
smaller than in equilibrium. In Fig. \ref{dels} we plot the entropy
difference per site, $\Delta s = s_{eq} - s$, versus $\rho$.  The
entropy reduction appears to attain a maximum near $\rho = 0.22$.
(For comparison we note that $R$ is maximum for $\rho \simeq 0.2$,
while the current in the driven lane is maximum for $\rho \simeq
0.17$.)  Under the nonequilibrium dynamics, curiously,
configurations with a more or less even distribution between the
lanes are favored.  This is demonstrated in Fig. \ref{sdn198}, which
shows the {\it average} probability over all configurations with
exactly $n$ particles in the driven lane, for the case $L=19$,
$N=8$.  (Of course, configurations with $n=N/2$ are more numerous,
but in equilibrium the average probability is the same for all $n$.)
The average probability is slightly higher for $n=3$ than for $n=5$,
consistent with the migration of particles to the undriven lane.
Note however that in the tails of the distribution ($n$=0, 1
compared with $n=$7, 8) this tendency is reversed.

A related question is whether the stationary probability
distribution $\overline{p}_i$ can be expressed in terms of a small
set of variables.  The `natural' set of variables characterizing the
global state of the system would appear to be the current densities
$j_d$ and $j_u$ and the density $\rho_d$ of particles in the driven
lane. Classifying configurations by the values of these three
variables, however, we find large variations in $\overline{p}_i$
within each class.  (In fact, for classes with more than one
element, the standard deviation of $\overline{p}_i$ over the class
is comparable to the mean stationary probability of the class.) For
example, in the case $L=16$, $n=8$, the most numerous class, with
1497 configurations, is that having $n_d = 4$, and total currents
$J_d = 4$ and $J_u = 0$.  Within this class the stationary
probabilities vary over two orders of magnitude (from $1.3 \times
10^{-6}$ to $1.23 \times 10^{-4}$), and almost all are distinct.
(There is just one pair of configurations that have the same
probability.) We therefore conclude that unlike in equilibrium, the
stationary probability distribution cannot be written as a function
of a reduced set of variables.

\section{Cluster approximations}

Numerical solution of the master equation, discussed in the previous
section, furnishes essentially exact results for limited system
sizes.  In this section we apply a complementary approach that
treats effectively infinite systems using dynamic cluster
approximations \cite{marro,mancam}.  We work with clusters of $2n$
sites (``$n$-column approximation").  For a given cluster size, we
enumerate all configurations and all possible transitions involving
the movement of a particle into, out of, or within the cluster. Next
we consider the master equation for the probability distribution on
this set of configurations. For certain transitions, the
contributions to the gain and loss terms in the master equation can
be written exactly using the $n$-column probability distribution. In
other cases, however, the probability of a larger configuration,
involving $n+1$ or $n+2$ columns, is needed to evaluate the
contribution.  Such probabilities are approximated in terms of the
$n$-column probabilities.

\begin{figure}[h]
\rotatebox{0}{\epsfig{file=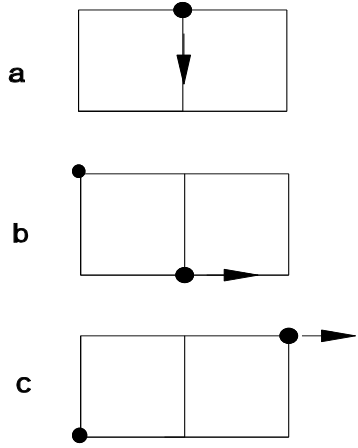,width=7.0cm,height=7.0cm}}
\caption{Examples of transitions considered in the cluster
approximation.} \label{moves}
\end{figure}

Consider, for example, the transitions depicted in Fig. \ref{moves}.
Using the notation defined in the previous section, the starting
configurations are denoted 6, 21, and 11, in cases a, b, and c,
respectively. We denote the probability of configuration $i$ on a
set of $n$ columns as $p_n(i)$.  In case a, the contribution to the
loss term in $d p_3(6)/dt$ [and to the corresponding gain term for
$p_3 (3)$] is simply $\beta p_3 (6)$, since this transition does not
depend on the state of any sites outside the 3-column cluster. In
case b, the transition is only possible if the site immediately to
the right of the target site is vacant.  This transition therefore
involves the 4-column probabilities $p_4(63)$ and $p_4(65)$.  Using
the factorization $p_4 (63) \simeq p_3(21) p_3 (9)/ p_2 (3)$ and the
analogous expression for $p_4 (65)$, the contribution (to the loss
term for $p_3 (21)$, in this case) due to this transition is

\begin{equation}
R_b = \gamma \frac{ p_3(21) [ p_3 (9) + p_3 (11)]}{p_2 (3)}.
\end{equation}

\noindent Similar reasoning leads, in case c, to the expression

\begin{equation}
R_c = \alpha \frac{ p_3(11) p_3 (6) [ p_3 (18) + p_3 (19)]}{p_2 (2)
p_2 (6)}.
\end{equation}
In general, in the $n$-column approximation, an ($n\!+\!1$)-column
probability is given by the product of the corresponding $n$-column
probabilities divided by the ($n\!-\!1$)-column probability
corresponding to the overlap region. (The latter is a marginal
probability obtained from the $n$-column probability distribution.)
Probabilities of ($n\!+\!2$)-column clusters are similarly estimated
as the product of three $n$-column probabilities divided by the
product of two ($n\!-\!1$)-column probabilities.

The systems of equations for 1, 2 and 3 columns are readily
constructed by hand, but to study larger clusters we develop a
computational algorithm that generates all possible transitions on
an $n$-column cluster, and applies the approximation scheme defined
above systematically.  The same routine integrates the resulting set
of coupled ODEs numerically using the fourth-order Runge-Kutta
scheme.  Due to relatively slow convergence, we limited this study
to $n \leq 7$ columns. The predictions of the cluster approximation
are compared against simulation in the following section.

To close this section we comment on a subtlety arising in the
definition of the initial probability distribution. Valid $n$-column
distributions must satisfy certain symmetry relations, due to the
fact that, in general, the marginal distributions for $m < n$
columns can be written in several ways. Consider the case $n=2$, for
which the possible configurations are 0, 1, 2, 3, 5, 6, and 7. In
addition to normalization, the following constraints apply to the
two-column probability distribution:

\begin{equation}
p_2(3) + p_2(5) = p_2 (1) + p_2(7) = p_1(1) = \rho_1
\end{equation}
and
\begin{equation}
p_2 (6) + p_2 (7) = p_2 (2) + p_2 (5) = p_1 (2) = \rho_2
\end{equation}

\noindent For $n=3$ further constraints arise, for example:
\begin{equation}
p_3 (2) + p_3 (20) + p_3 (11) = p_3(6) + p_3(7) = p_2(2)
\end{equation}
and
\begin{equation}
p_3(15) + p_3(16) = p_3(5) + p_3(23) = p_2(5)
\end{equation}

Since the total density $\rho$ is conserved by the equations of
motion, it must be defined in the initial distribution: $\rho =
(1/2L) \sum_j N_j p_n (j,t=0)$, where $N_j$ denotes the number of
occupied sites in configuration $j$.  For small clusters it is
possible to find a simple parametrization of the probability
distribution that permits one to fix the density while satisfying
the symmetry relations noted above.  But for large $n$ this becomes
increasingly more complicated.  To circumvent this difficulty, we
generate the initial distribution for the ladder by integrating the
equations of motion for the probability distribution (in the
$n$-column approximation), for a {\it different process}, namely,
the equilibrium lattice gas with nearest-neighbor exclusion,
evolving via single-particle adsorption and desorption. In this
process the fugacity $z$ controls the equilibrium density. We start
with density zero; at any subsequent moment the probability
distribution satisfies the conditions noted above, since they are
intrinsic symmetries of the equations of motion.  (In practice we
halt the integration when the density $\rho$ has reached the value
we wish to study in the driven model. There is no need to run the
adsorption-desorption model to equilibrium.)

\section{Simulation results}

\subsection{Lattice gas model}

\begin{figure}[h]
\rotatebox{0}{\epsfig{file=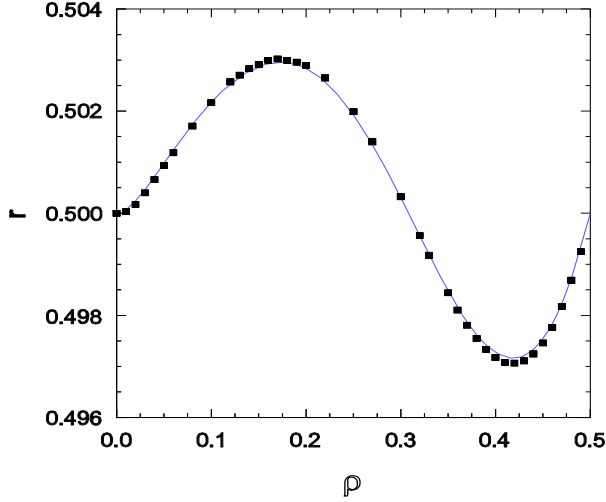,width=7.0cm,height=7.0cm}}
\caption{Stationary fraction $R$ of particles in the undriven lane
versus total density $\rho$, in simulations with $L=200$ (points)
compared with the seven-column cluster approximation (curve).}
\label{dr2}
\end{figure}

\begin{figure}[h]
\rotatebox{0}{\epsfig{file=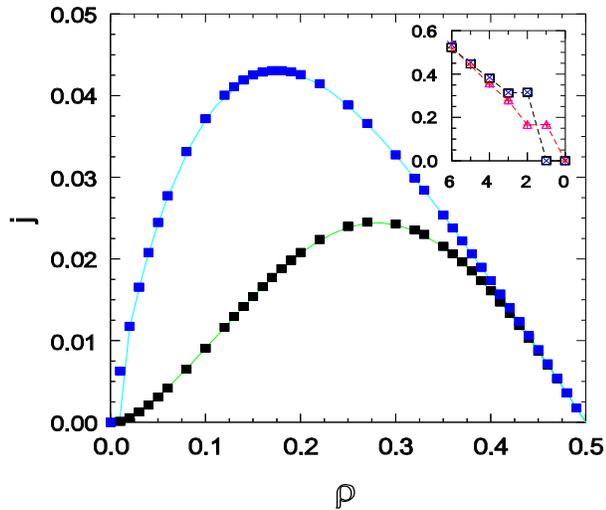,width=7.0cm,height=7.0cm}}
\caption{Stationary current densities $j_d$ (upper set) and $j_u$
(lower set) versus total density $\rho$, in simulations with $L=200$
(points) and the seven-column cluster approximation (curve). Inset:
total current versus number of holes, for $L=50$ (squares: undriven
lane, $\times$: driven) and $L=51$ (triangles: undriven, $+$:
driven); dashed lines are to guide the eye.} \label{drcurr}
\end{figure}

We perform Monte Carlo simulations of the lattice gas model defined
in Sec. II on ladders of $2 \times L$ sites with periodic
boundaries, using $L=200$ and 1000.  The results reported here
represent an average over 4 to 10 independent realizations.  In each
case, we verified that the system had reached the steady state. In
Fig. \ref{dr2} we show the density ratio $R = \rho_1/\rho$ for
transition rates $\alpha = 2/3$, $\beta = \gamma = 1/3$; its
behavior is very similar to that observed in the stationary solution
of the master equation, Fig. \ref{sd1619r}. Excellent agreement
between simulation and the cluster approximation (shown here for
$n=7$) is also evident. Similar observations apply to the currents,
depicted in Fig. \ref{drcurr}. [It is worth noting that small
cluster approximations do not provide good predictions. For example,
it is only for $n \geq 4$ that the approximation predicts the
migration to the driven lane ($R < 1/2$), observed at higher
densities.] Studies using $L=1000$ reveal little difference from the
results for $L=200$ shown in Figs. \ref{dr2} and \ref{drcurr}, so
that finite-size effects appear to be quite weak in this system.  In
fact, $R$ is only slightly less than found in the studies of much
smaller systems (Fig. \ref{sd1619r}). The current densities are also
comparable if one recalls that the rates used in the simulations are
1/3 those used in the master equation studies. The odd-even effects
noted in the latter analysis reappear here, as shown in the inset of
Fig. \ref{drcurr}, where the total current in each lane is plotted
versus the number of holes, $n_h \equiv L-N$. The currents are quite
different for $n_h \leq 4$.

Varying the hopping rates, the magnitude of the density shift
changes.  Fixing $\rho = 0.17$, the value for which $R$ is maximum
in Fig. \ref{dr2}, we obtain the largest value, $R = 0.5065$, for
$\beta/\alpha \geq 10$ and $\alpha/\gamma \geq 10$.  In the opposite
limit, $\gamma/\alpha \geq 10$, we observe very little migration
($|R - 1/2| \simeq 10^{-4}$), regardless of the value of the
interlane transition rate $\beta$.  The density imbalance also
disappears when we remove excluded-volume interactions between
particles in different lanes.

\begin{figure}[h]
\rotatebox{0}{\epsfig{file=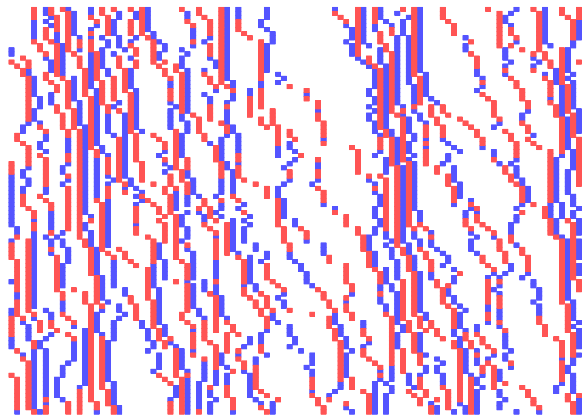,width=7.0cm,height=7.0cm}}
\caption{Space-time evolution, as explained in text. Rates $\alpha =
2/3$, $\beta = \gamma = 1/3$, density $\rho = 0.17$.} \label{cp21}
\end{figure}

Some insight into the dynamics is afforded by a space-time plot of
the evolution.  Fig. \ref{cp21} shows the configuration at unit time
intervals, in a system that has already relaxed to the stationary
state. (Time increases downward in the figure.  Particles in
different lanes are given different colors.)  There are rarefied,
high-mobility regions separated by ``bottlenecks" that typically
form behind a diagonal pair of particles, with the one to the right
in the undriven lane.

\begin{figure}[h]
\rotatebox{0}{\epsfig{file=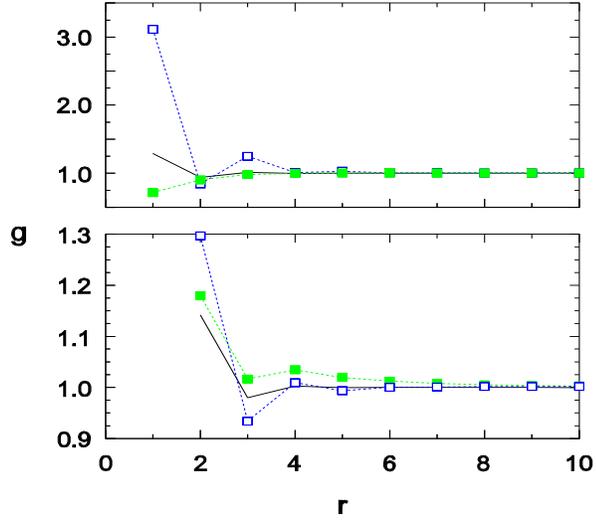,width=7.0cm,height=7.0cm}}
\caption{Radial distribution functions, density $\rho=0.125$,
$L=200$. Lower panel: filled symbols: $g_{uu}$: open symbols:
$g_{dd}$, solid line: $g_s$ (equilibrium). Upper panel: filled
symbols: $g_{ud}$: open symbols: $g_{du}$, solid line: $g_d$
(equilibrium).} \label{g125ab}
\end{figure}

Correlations in the driven system are quite different from those in
equilibrium.  In this model we may distinguish four radial
distribution functions (RDFs): $g_{uu}$, $g_{dd}$, $g_{ud}$ and
$g_{du}$.  In each case, $g_{mn} (r)$ is proportional to the
probability of sites $i$ and $i+r$ (in lanes $m$ and $n$, resp.),
being occupied simultaneously, and is normalized so that $g_{mn}(r)
\to 1$ for $r \to \infty$.  In Fig. \ref{g125ab} we compare the RDFs
with the corresponding equilibrium distributions, for density $\rho
= 0.125$.  (In equilibrium we need only distinguish $g_s$, for sites
in the same lane, and $g_d$, for sites in different lanes.)
Correlations are generally stronger in the driven system than in
equilibrium, especially for $g_{du}$ at contact, reflecting the
accumulation of driven particles just behind undriven ones.  It is
also evident that $g_{uu}$ decays more slowly than the corresponding
equilibrium RDF.

\begin{figure}[h]
\rotatebox{0}{\epsfig{file=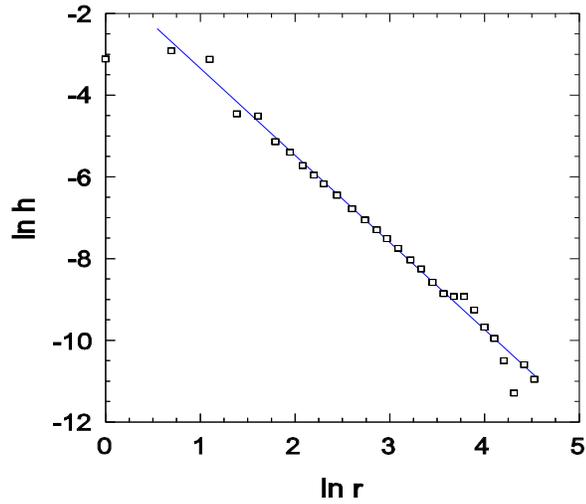,width=7.0cm,height=7.0cm}}
\caption{Correlation function $h(r)$ for density $\rho = 0.125$. The
slope of the line is -2.11.} \label{lh125}
\end{figure}

The slow decay is in fact shared by all of the RDFs in the driven
system; each appears to follow a power law.  In Fig. \ref{lh125} we
plot the correlation function $h(r) = g(r) - 1$ on log scales, where
$g(r)$ denotes the average of the four RDFs defined above.  The
correlation function follows $h(r) \propto r^{-\phi}$ with $\phi =
2.11(3)$ for density $\rho = 0.125$.  We observe similar power-law
tails in $h(r)$ at other densities (up to $\rho = 0.3$), with decay
exponents $\phi \approx 2$, as summarized in Table I.  (These
results were obtained using systems with $L \geq 500$ sites.) At
higher densities the RDF exhibits damped oscillations with an
exponentially decaying amplitude: $|h(r)| \propto e^{-r/\xi}$, as in
equilibrium. (For $\rho=0.4$, for example, we find $\xi \simeq
3.4$.) In fact, the RDFs at high density differ from their
equilibrium counterparts by only a few percent. The power-law tail
may still be present at high density, but due to its low amplitude
it tends to be masked by the oscillations (for smaller $r$) and
noise (at larger $r$).

\begin{center}
%\caption
\begin{tabular}{c c}
\hline \hline
 $\rho$ & $\phi$    \\
\hline
 0.05    & 1.93(4)  \\

 0.1     & 1.98(2)  \\

 0.125   & 2.10(3)  \\

 0.15    & 1.96(2)   \\

 0.20    & 1.99(3)   \\

 0.25    & 1.95(5)   \\

 0.30    & 2.16(13)  \\
\hline \hline
\end{tabular}
\end{center}
\vspace{1em}

 {\sf Table I. Values of the exponent $\phi$ governing
the decay of the two-point correlation function in the lattice model
with $\alpha=2/3$ and $\beta=\gamma=1/3$.} \vspace{2em}

As the particle density tends to its maximum value of 1/2, the
density of {\it mobile} vacancies $\rho_m$ approaches zero linearly
[$\propto (1/2 -\rho)$], as does that of mobile particles.  For a
vacancy to be mobile, it must be capable of exchanging places with a
neighboring particle, which means that the vacancy must have exactly
one occupied nearest neighbor.  In the undriven lane, this can be
any of the three neighboring sites, but for a vacancy in the driven
lane, the neighboring particle must be below or to the left, for it
to be mobile.  Thus at low density we expect the density
$\rho_{m,d}$ of mobile vacancies in the driven lane to be 2/3 that
in the undriven lane, as is verified in simulations. Surprisingly,
at high density the disparity becomes much greater, the ratio
$\rho_{m,d}/\rho_{m,u}$ tending to 1/4 as $\rho \to 1/2$. At the
same time, there is a marked tendency for mobile vacancies to
cluster in the driven layer; the contact value of the associated RDF
is $\simeq 8$ at density 0.49.

\subsection{Continuous-space model}

\begin{figure}[h]
\rotatebox{0}{\epsfig{file=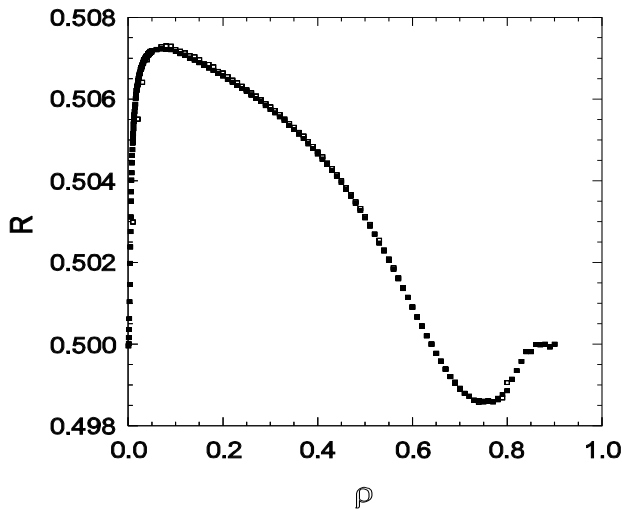,width=7.0cm,height=7.0cm}}
\vspace{1cm}

\caption{Stationary fraction $R$ of particles in the undriven lane
versus total density $\rho$, in simulations of the continuous-space
model with rates $\alpha=0.8$, $\beta=1$, and $\gamma = 0.1$.  Open
squares: $L=200$; filled squares: $L=1000$.} \label{dr2csr}
\end{figure}

\begin{figure}[h]
\rotatebox{0}{\epsfig{file=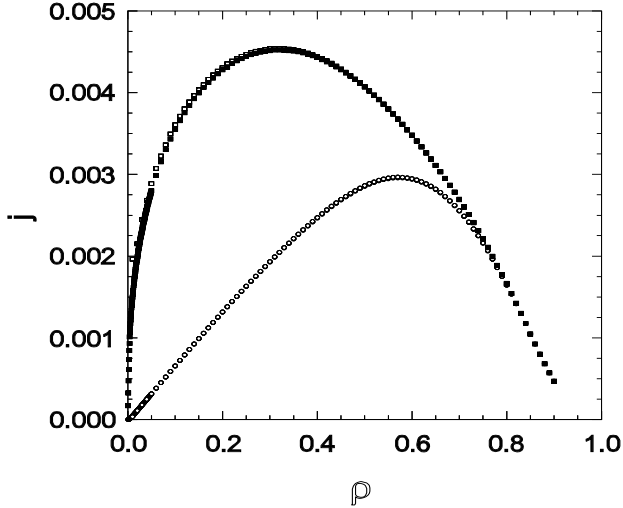,width=7.0cm,height=7.0cm}}
\vspace{1cm}

\caption{Stationary current densities $j_d$ (upper set) and $j_u$
(lower set) versus total density $\rho$, in in simulations of the
continuous-space model, parameters as in Fig. \ref{dr2csr}.}
\label{dr2csj}
\end{figure}

We also performed simulations of a system in which, as described in
Sec. II, the $x$-coordinate of a particle can vary continuously on
the interval $[0, L]$, again with periodic boundaries.  A
preliminary investigation showed that migration to the undriven lane
is strongest (for densities near 0.17) for rates $\alpha=0.8$,
$\beta=1$, and $\gamma = 0.1$. Using these values, we study the
dependence of the lane densities  and currents as a function of
overall density $\rho$, in systems of size $L=200$ and 1000.  The
fraction $R$ of particles in the undriven lane (Fig. \ref{dr2csr})
shows the same general tendencies as in the lattice gas, but the
initial rise with density $\rho$ is more rapid, and the reverse
migration at high densities is weaker than on the lattice. The
current densities (Fig. \ref{dr2csj}) are again qualitatively
similar to those found in the lattice model.

\begin{figure}[h]
\rotatebox{0}{\epsfig{file=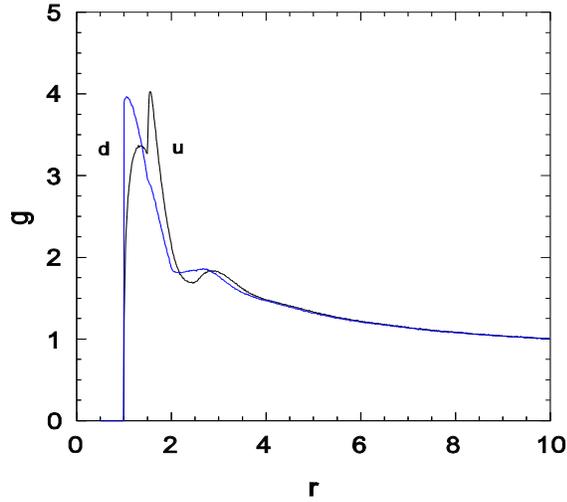,width=7.0cm,height=7.0cm}}
\caption{Stationary radial distribution function $g(r)$ in the
continuous space model, $\rho = 0.1$, $L=200$, for particles in the
same lane; d: driven lane; u: undriven lane.} \label{grslnew}
\end{figure}

\begin{figure}[h]
\rotatebox{0}{\epsfig{file=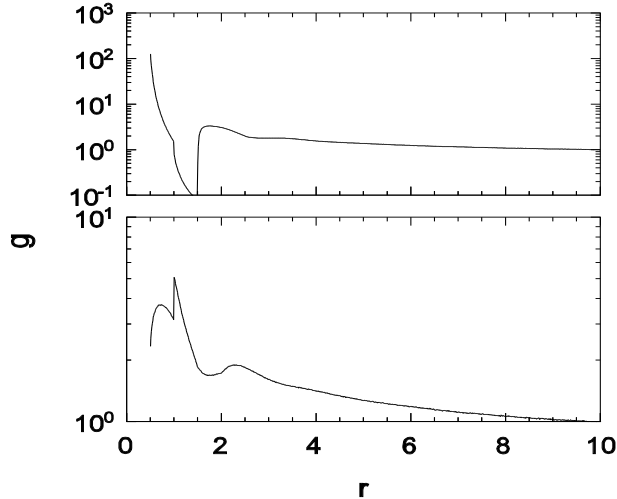,width=7.0cm,height=7.0cm}}
\caption{Stationary radial distribution function $g(r)$ as in Fig.
\ref{grslnew}, for particles in different lanes; upper panel:
$g_{du}$; lower: $g_{ud}$.} \label{grdlnew}
\end{figure}

Examples of RDFs in the continuous-space model are plotted in Figs.
\ref{grslnew} and \ref{grdlnew} for density $\rho = 0.1$; they
exhibit the familiar peak near contact and damped oscillations. The
value at contact however is much larger than in equilibrium, where
for the Tonks gas $g(1) = 1/(1-\rho)$. In particular, $g_{du}$ (Fig.
\ref{grdlnew}) becomes extremely large as $r \to 0.5$, reflecting
the pile-up of driven particles behind undriven ones in the opposite
lane. This feature allows to understand why the RDF in the undriven
lane attains its maximum near $r=1.5$ rather than $r=1$. Given the
exceptionally large value of $g_{du}$ near $r=0.5$, we expect (and
in fact observe) a secondary maximum in this function near $r=1.5$,
due to situations in which particles $i-1$ and $i$ are in the driven
lane and $i+1$ is in the undriven one. The maximum in $g_{uu}$ near
$r=1.5$ is due to events in which particle $i-1$ jumps to the
undriven lane.

\begin{figure}[h]
\rotatebox{0}{\epsfig{file=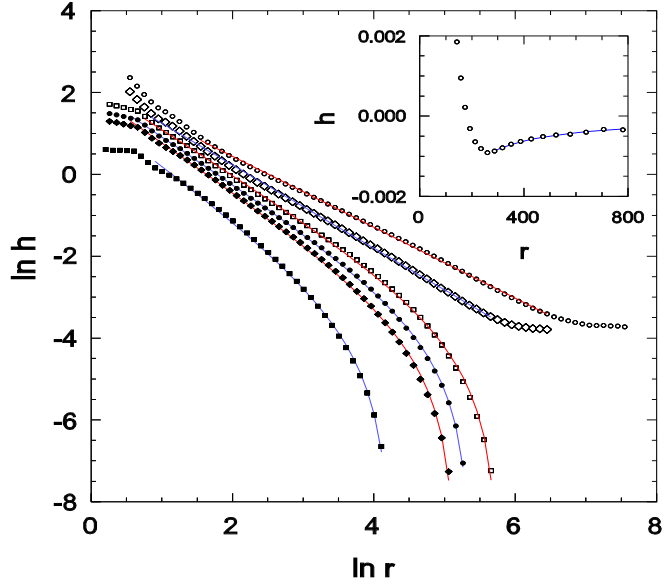,width=7.0cm,height=7.0cm}}
\vspace{1cm}

\caption{Correlation function $h(r)$ for (upper to lower) $\rho =
0.01$, 0.02, 0.03, 0.04, 0.05, and 0.10. Lines represent fitting
functions: a simple power law for $\rho \leq 0.02$, and the function
$H(r)$ defined in Eq. \ref{hfit} for $\rho \geq 0.03$.  Inset:
detail of $h(r)$ for large $r$ on linear scales; fitting function:
-$0.25/r$.} \label{f16}
\end{figure}

To obtain precise results for the correlation functions $h_{\alpha
\beta} = g_{\alpha \beta} - 1$ at large $r$ we performed studies
using larger systems (5000 - 20$\,$000 sites) and improved
statistics (6 - 10 independent runs consisting of $\geq 2 \times
10^8$ time steps). (Since all four correlation functions show the
same behavior for large $r$, we study the mean of the four, denoted
simply as $h(r)$.) At the lowest densities studied ($\rho = 0.01$
and 0.02, using $L=20\,000$) we find a clear power-law decay over
more than two decades (see Fig. \ref{f16}). Comparing Figs.
\ref{f16} and \ref{lh125}, one sees that in continuous space, the
amplitude of the power law is nearly two orders of magnitude larger
than on the lattice. For $\rho \geq 0.03$, the correlation function
$h(r)$ decays in a nonmonotonic fashion: it is negative for $r>r^*$,
and approaches zero from below as $r \to \infty$ (see Fig. \ref{f16}
inset). In these cases we cannot fit $h(r)$ with a simple power law,
but have obtained excellent fits (for $r < r^*$), introducing one
further adjustable parameter, in the functional form

\begin{equation}
H(r) \simeq A r^{-\phi} \left( 1 - \frac{r}{r^*} \right) ^\psi
\label{hfit}
\end{equation}
(Note that $r^*$ is determined from the simulation data and is not
an adjustable parameter.  $\phi$ and $\psi$ are obtained by
minimizing the variance of $\ln h(r) - \ln H(r)$, ignoring the
additive constant $A$.  The latter is obtained once the two
exponents have been determined.)  The simulation data and associated
fits are shown in Fig. \ref{f16}; the parameters, listed in Table
II, suggest that the decay exponent $\phi$ varies systematically
with density. (We cannot fit the region of negative $h(r)$ using
this function; the data for $\rho = 0.05$  shown in the inset of
Fig. \ref{f16}, suggest that $|h(r)| \sim 1/r$ in this regime.)

Preliminary results indicate that $\phi$ also depends on the
transition rates. Fixing the density at $\rho = 0.05$, the rates
$\alpha=0.8$, $\beta=1$, and system size $L=1000$, we find $\phi =
1.21(7)$, 1.15(2), and 1.51(2) for $\gamma = 0.02$, 0.1, and 0.5,
respectively. For density $\rho=0.25$ the correlation functions have
an oscillatory structure; the initial decay is exponential, with an
associated correlation length of about 3.2. We verify that when {\it
both} lanes are driven, the correlation function decays
exponentially, as expected.

\vspace{2em}

\newpage

\begin{center}
%\caption
\begin{tabular}{c c c c}
\hline \hline
 $\rho$ & $\phi$ & $\psi$  & $r^*$    \\
\hline
 0.01    & 0.87(1) & - & - \\

 0.02    & 1.02(1) & - & - \\

 0.03    & 1.10(3) & 1.47(3) & 317.3 \\

 0.04    & 1.14(2) & 1.37(3) & 212.7 \\

 0.05    & 1.19(3) & 1.44(3) & 173.3 \\

 0.1     & 1.28(3) & 1.43(3) &  68.7 \\

 0.15    & 1.49(4) & 1.36(4) &  45.5 \\
\hline \hline
\end{tabular}
\end{center}
\vspace{1em}

 {\sf Table II. Exponent $\phi$ and related parameters
characterizing the two-point correlation function in the
continuous-space model. Transition rates $\alpha=0.8$, $\beta=1$,
and $\gamma=0.1$, system size $L \geq 5000$.} \vspace{2em}

%\vspace{2em}

\section{Redistribution mechanism}

The cluster approximation developed in Sec. IV reproduces the
simulation results to high accuracy, but does not explain the
observed migration of particles. In this section we suggest a
mechanism for this phenomenon, based on the tendency (as revealed in
the correlation functions discussed above) for diagonal pairs of
particles to form, with the particle to the left in the driven lane
[that is, pairs of the form $(x,2)$, $(x\!+\!1,1)$].  Given that
such pairs are relatively long-lived, it is interesting analyze the
problem of a third particle in the presence of such a pair.

Suppose the fixed pair occupies sites (0,2) and (1,1) on a ring of
$2L$ sites. Then the third particle is free to visit sites $(3,1),
(4,1), ..., (L\!-\!1,1)$ and $(2,2), (3,2), ..., (L\!-\!2,2)$. In
the stationary state the third particle will tend to be found close
to, and to the left of, the fixed pair, and the stationary
probability $\overline{p}(L\!-\!j,y)$ should decay exponentially
with $j$ due to the rightward drift.  It turns out that in this
situation, the third particle has a higher probability to be found
in the undriven lane than in the driven one, as we now show.

From here on we refer to the third particle simply as the
`particle', and analyze the master equation describing this biased
random walk, which is subject to reflecting boundaries, due to the
fixed pair. Denoting by $p(x,y,t)$ the probability for the particle
to be at site $(x,y)$ at time $t$, we have, away from the
boundaries,

\begin{equation}
\dot{p}(x,1) = -(2 \gamma + \beta)p(x,1) + \gamma[p(x-1,1) +
p(x+1,1)] + \beta p(x,2) \label{dpx1dt}
\end{equation}
and
\begin{equation}
\dot{p}(x,2) = -(\alpha + \beta)p(x,2) + \alpha p(x-1,2) + \beta
p(x,1) \label{dpx2dt}
\end{equation}
while at the right boundary the equations of motion are
\begin{equation}
\dot{p}(L\!-1,1) = -\gamma [p(L\!-\!1,1) - p(L\!-\!2,1)]
\label{dpMm1dt}
\end{equation}
and
\begin{equation}
\dot{p}(L\!-\!2,2) = -\beta[p(L\!-\!2,2) - p(L\!-\!2,1)] + \alpha
p(L\!-\!3,1)  \label{dpMm2dt}
\end{equation}
There is also a reflecting boundary at (2,1) [state (1,2) is
transient], but in the limit of large $L$ the probability in this
neighborhood is negligible since the solution decays exponentially
with distance from the right boundary: $p(L\!-\!z,y) \propto
\exp[-Const. \times z]$.  It is straightforward to verify that in
this limit the stationary solution is:

\begin{equation}
p(L\!-\!2,1) = p(L\!-\!1,1) = C
\end{equation}
\begin{equation}
p(L\!-\!j,1) = \kappa^{(j-2)} C ,\;\;\;\;\;\; (j \geq 2)
\end{equation}
\begin{equation}
p(L\!-\!2,2) = C \left[ 1 + \frac{\gamma}{\beta} (1 - \kappa)
\right]
\end{equation}
\begin{equation}
p(L\!-\!j,2) =  \kappa^{(j-3)} \tilde{\gamma} (1 - \kappa)  C,
\;\;\;\;\;\; (j \geq 3)
\end{equation}

where

\begin{equation}
C^{-1} = 1 + \tilde{\gamma} + \frac{\gamma}{\beta} (1-\kappa) +
\frac{2 - \kappa}{1-\kappa} \label{Crecip}
\end{equation}
is a normalization factor, and

\begin{equation}
\kappa = 1 + \frac{\eta}{2}(1 + \tilde{\gamma}) - \sqrt{\eta +
\frac{\eta^2}{4}(1+\tilde{\gamma})^2} \label{kappa}
\end{equation}
with $\eta = \beta/\gamma$.  Since $\kappa < 1$ for any $\eta > 0$,
the stationary probability distribution decays exponentially with
distance from the boundary.  In the stationary state, the
probability to find the particle in the undriven lane is

\begin{equation}
p_1 = \sum_{j=1}^\infty p(L\!-\!j,1) = \frac{1}{1+
\frac{1-\kappa}{2-\kappa} \left[ 1 + \tilde{\gamma}
+\frac{\gamma}{\beta} (1-\kappa)\right]} \label{pyeq1}
\end{equation}
\vspace{1em}

\noindent In general, $p_1$ is greater than 1/2, demonstrating the
tendency for the particle to migrate to the undriven lane; this
probability varies between 1/2, for $\gamma \ll \beta$, and 2/3, in
the opposite limit.   [This despite the fact that $p(x,y)$ takes its
maximum in the driven lane, at $(L\!-\!2,2)$.] For fixed $\beta$ and
$\gamma$, $p_1$ increases with $\alpha$. These tendencies are in
qualitative agreement with the trends regarding particle migration
noted in simulations.  We note in passing that as $L \to \infty$,
$p_1$ approaches the limiting value, Eq. (\ref{pyeq1}), from above.

We have not found a simple argument to explain the tendency for
particles to accumulate in the {\it driven} lane at high densities.
(Due to lack of particle-hole symmetry, we cannot apply the
low-density argument given above to the dynamics of holes at high
density.)  We determined the radial distribution functions for pairs
of holes $\overline{g}_{ij}$, defined in analogy to the particle
RDFs discussed above. At high density these functions show marked
oscillations (due to excluded volume interactions), and decay
exponentially due to the presence of uncorrelated vacancies.  We
find that the probability of a diagonal pair of holes is higher if
the hole on the right is in the driven lane.  (The relative
difference between $\overline{g}_{du}$ and $\overline{g}_{du}$ is
about 15\% for density 0.45.)  By the argument given above, this
asymmetry suggests a tendency for vacancies to migrate to the
undriven lane, hence a higher particle density in the driven one.

\section{Discussion}

We study nonequilibrium stationary states in a lattice gas with
nearest-neighbor exclusion on a periodic ladder, in which particles
in one lane are driven, while in the other lane the dynamics is
unbiased.  The interlane hopping rates are symmetric.  Numerical
solution of the master equation, cluster approximations, and Monte
Carlo simulation all reveal a tendency for particles to migrate to
one of the lanes. At low densities particles accumulate in the
undriven lane, while at higher densities the driven lane is favored.
Similar results are found in simulations of a continuous space
model. Redistribution of particles is strongest when the hopping
rate in the driven lane is considerably larger (by a factor of ten
or so) than the rate in the undriven lane.

The redistribution found here is qualitatively similar to that
observed in the two-dimensional driven NNE lattice gas with a linear
drive profile \cite{nne-shear}, but here the effects are much
weaker, and migration to the driven lane is not accompanied by
jamming, as it is in the two-dimensional system. The density
difference observed in the one-dimensional system is relatively
small, on the order of 1\% or less of the total density; in the
two-dimensional system this difference can reach 25\%. It is
possible that migration in the two-dimensional system represents an
amplification of the effect observed here, as particles move
progressively to regions with weaker drive. (On the other hand the
drive gradient in the two-dimensional system is $\propto 1/L$, that
is, much weaker than in the ladder studied here.)

Independent of its connection with the two-dimensional case, the
redistribution observed here is an interesting nonequilibrium effect
in a simple driven system. The existence of distinct densities is
not in itself surprising: since the dynamics in the two lanes are
radically different, there is no reason for the associated particle
densities to be equal.  But this does not explain why particles
migrate to the undriven lane at low density, and to the driven one
at high density.  The former effect can be understood noting the
high probability of diagonal pairs of particles, with the particle
to the left in the driven lane, and the effect such a pair has on a
third particle.

We have generated the stationary probability distribution on rings
of up to 2$\times$19 sites.  The stationary particle and current
densities are quite similar to those for larger systems (obtained
via Monte Carlo simulation).  The results for the stationary
probability distribution allow us to characterize the reduction in
statistical entropy caused by the drive. They also lead to the
conclusion that the stationary probability distribution cannot be
written in terms of a reduced set of variables.  In general, each
configuration has a distinct stationary probability.

Remarkably, the two-point correlation function, $h(r)$, decays as a
power law at low densities.  The decay exponent $\phi$ takes a value
close to two in the lattice gas. In the off-lattice system, $\phi$,
takes a value near or below unity at low densities, and increases
continuously with density; the exponent also varies as a function of
the transition rates.  The amplitude of the power-law is much larger
in the off-lattice system. At higher densities $h(r)$ exhibits the
oscillatory structure familiar in dense fluids, with an envelope
that decreases exponentially with distance. While algebraic decay of
correlations is expected to be a generic feature of driven systems
with conserved density \cite{garrido90}, the existing analyses do
not apply specifically to the system studied here, and do not
predict an exponent that varies continuously as a function of the
density.

A precedent for algebraic decay of correlations in a periodic
one-dimensional system may be found in directed percolation
\cite{marro}, and in the driven lattice gas studied by Helbing et
al. \cite{helbing99}.  In these examples, however, slow decay of
correlations is a manifestation of a continuous phase transition.
The system studied here shows now evidence of a phase transition
(the density remains uniform, and there is no hint of nonanalytic
behavior in the current, for example), and so appears to possess
only a disordered phase.  Indeed, algebraic decay of correlations is
expected to occur in the {\it high-temperature} phase of driven
systems \cite{garrido90}.  A natural question is why, at higher
densities ($\rho \geq 0.25$ in the continuous-space model),
correlations appear to decay exponentially.  We suspect that $h(r)$
continues to have possess a power-law tail, but that this feature is
masked by the short-range oscillatory contribution, due to its much
larger amplitude.  Verification of this conjecture will have to
await more extensive simulations or theoretical analysis.

Several one-dimensional driven models have been solved exactly via
the matrix-product method \cite{derrida,krebs,klauck}. While it is
possible that the model studied here could be solved by this method,
we note that it corresponds, in the scheme of Ref. \cite{klauck}, to
a model with $m=3$ states per site (a site here corresponding to a
column in our original description), with an `internal transition'
between two of these states, and an interaction range of $r=3$. Thus
the search for a set of matrices satisfying the appropriate algebra
promises to be a complex task, which we defer to future work.

The present study leaves several other questions open for future
theoretical and numerical investigation.  Principal amongst these is
an explanation of the algebraic decay of correlations, and a more
precise determination of the exponent $\phi$ as a function of
density and transition rates.  Study of higher-order correlations,
and of time-dependent properties, would also be of interest.
\vspace{2em}

%\newpage
\noindent{\bf Acknowledgments}

We are grateful to Royce Zia and Gunter Sch\"utz for valuable
comments. This work was supported by CNPq and FAPEMIG, Brazil.

\newpage

\bibliographystyle{apsrev}

\end{document}